\documentclass[twocolumn,english,aps,pra,showpacs,superscriptaddress,amsmath,amsfonts]{revtex4-1}

\usepackage{epsfig}
\psfigdriver{dvips}
\usepackage{graphicx}
\usepackage{color}

\usepackage{bm, bbm}      
\usepackage{curves}
\usepackage{epic}
\usepackage{wasysym}
\usepackage{subfigure}

\usepackage[english]{babel}

\begin{document}

\title{ Hole propagation in the Kitaev-Heisenberg model:\\
From quasiparticles in quantum N\'eel states to non-Fermi liquid in the Kitaev phase}

\author{Fabien Trousselet}
\affiliation{Max-Planck-Institut f\"ur Festk\"orperforschung,
             Heisenbergstrasse 1, D-70569 Stuttgart, Germany}
\affiliation{University Grenoble Alpes, Institute N\'eel, 25 Avenue des Martyrs, BP166,
             F-38042 Grenoble Cedex 9, France}
\affiliation{CNRS, Institute N\'eel, 25 Avenue des Martyrs, BP166,
             F-38042 Grenoble Cedex 9, France}

\author{Peter Horsch}
\affiliation{Max-Planck-Institut f\"ur Festk\"orperforschung,
             Heisenbergstrasse 1, D-70569 Stuttgart, Germany}

\author{Andrzej M. Ole\'s}
\affiliation{Max-Planck-Institut f\"ur Festk\"orperforschung,
             Heisenbergstrasse 1, D-70569 Stuttgart, Germany}
\affiliation{Marian Smoluchowski Institute of Physics,
             Jagellonian University, Reymonta 4, PL-30059 Krak\'ow, Poland}

\author{Wen-Long You}
\affiliation{Max-Planck-Institut f\"ur Festk\"orperforschung,
             Heisenbergstrasse 1, D-70569 Stuttgart, Germany}
\affiliation{School of Physical Science and Technology, Soochow University,
             Suzhou, Jiangsu 215006, People's Republic of China }

\date{\today}

\begin{abstract}
We explore with exact diagonalization the propagation of a single hole
in four magnetic phases of the $t$-$J$-like Kitaev-Heisenberg model on 
a honeycomb lattice: the N\'eel antiferromagnetic, stripe, zigzag and
Kitaev spin-liquid phase. We find coherent propagation of spin-polaron 
quasiparticles in the antiferromagnetic phase by a similar mechanism as
in the $t$-$J$ model for high-$T_c$ cuprates.
In the stripe and zigzag phases clear quasiparticles features appear
in spectral functions of those propagators where holes are created and 
annihilated on one sublattice, while they remain largely {\it hidden} 
in those spectral functions that correspond to photoemission experiments. 
As the most surprising result, we find a totally incoherent spectral 
weight distribution for the spectral function of a hole moving in the 
Kitaev spin-liquid phase in the strong coupling regime relevant for 
iridates. At intermediate coupling the finite systems calculation 
reveals a well defined quasiparticle at the $\Gamma$ point, however, 
we find that the gapless spin excitations wipe out quasiparticles at 
finite momenta. Also for this more subtle case we conclude that in the 
thermodynamic limit the lightly doped Kitaev liquid phase does not 
support quasiparticle states in the neighborhood of $\Gamma$, and 
therefore yields a {\it non-Fermi liquid}, contrary to earlier 
suggestions based on slave-boson studies.
These observations are supported by the presented study of the 
dynamic spin-structure factor for the Kitaev spin liquid regime.
\end{abstract}

\pacs{75.10.Kt, 05.30.Rt, 71.10.Hf, 79.60.-i}

\maketitle

\section{Introduction}

Carrier propagation in Mott or charge-transfer insulators is a
challenging problem particularly motivated by strongly correlated
superconducting cuprates \cite{Dag94,Szc90,Eme97,Lee06}. While holes 
move incoherently in one-dimensional systems featuring charge and spin 
separation \cite{And90} as well as in systems with antiferromagnetic 
(AF) Ising interactions \cite{Bri70,Tru88}, the hole motion becomes 
coherent and quasiparticles (QPs) arise at low energy
in the quantum AF $t$-$J$ model \cite{Mar91}.
These QPs are indeed observed in angle-resolved photoemission (ARPES)
experiments in cuprates \cite{And03}. In general, low-energy QPs coexist
with incoherent processes at high energy, as in the AF phase on the
square \cite{Mar91} or honeycomb lattice \cite{Sus06}. This is however 
not always the case as shown by ARPES experiments for the spin-orbit 
Mott insulator Na$_2$IrO$_3$ \cite{Com12}, without clear evidence for 
QPs at low energy \cite{Tro13}.
In a recent study hole-doped Li$_2$Ir$_{1-x}$Ru$_x$O$_3$ with honeycomb
structure was found insulating at all doping levels \cite{Lei13}.

Electronic systems with honeycomb lattice include graphene \cite{Gra09,Pol13},
optical lattices \cite{Wun08}, topological insulators \cite{Hoh13}, and
frustrated magnets \cite{Nor09,Bal10,Kit06,Nus13,Jac09,Mur10,Cha10,Tre11}.
Interest in the latter was triggered by theoretical predictions
\cite{Jac09,Shi09} that Na$_2$IrO$_3$ may host Kitaev model physics and
quantum spin Hall effect. The ground state of this model is a Kitaev
spin liquid (KSL) characterized by finite spin correlations only for
nearest neighbor (NN) spins \cite{Bas07}. 
The KSL belongs to the spin disordered phases which are in the center of 
interest in quantum magnetism \cite{Nor09,Bal10}.
In Mott insulators where the strong spin-orbit interaction generates a 
Kramers doublet from partly filled $t_{2g}$ orbitals \cite{Jac09,Cha10}, 
as in Na$_2$IrO$_3$, effective $S=1/2$ pseudospins stand for local, 
spin-orbital entangled $t_{2g}$ states which form orbital moments 
\cite{Hor03}. Both Kitaev and Heisenberg interactions emerge from the 
spin and orbital coupling on the honeycomb lattice of iridium ions
and form the Kitaev-Heisenberg (KH) model
\cite{Jac09}. Experimental observations revealed zigzag (ZZ) magnetic
order in Na$_2$IrO$_3$ \cite{Liu11,Sin12,Ye12} --- it may be explained
within the KH model with next-nearest neighbor (NNN) $J_2$ and third NN
(3NN) $J_3$ AF interactions \cite{Cha10,Tre11}. The phase diagram of
the frustrated Heisenberg $J_1$-$J_2$-$J_3$ AF model \cite{Alb11,Gan13}
includes the N\'eel AF, ZZ, and also stripy (ST) phase.
These phases survive when more general spin interactions with 
symmetric off-diagonal exchange are considered \cite{Rau14}.

Doping the KSL is particularly exciting as the ground state is spin
disordered and it is unclear whether QPs would form \cite{Sen03}.
Moreover in the KSL interactions may emerge that lead to
unexpected forms of superconductivity. Indeed, slave-boson studies 
found here $p$-wave superconductivity at intermediate doping 
\cite{Hya12,You12,Oka13,Bur11}, whereas Fermi liquid was claimed at 
light doping \cite{Mei12}. An important first step to explore hole
doping is however the study of single hole motion and the existence of
QPs. So far, spin-charge separation was shown for the kagome lattice,
being a prototype of a spin liquid, while small QP peaks were found at
some momenta in a frustrated checkerboard lattice \cite{Lau04}.
The $t$-$J$-like KH model provides here a unique opportunity to
investigate hole propagation in quite different magnetic
phases emerging from frustrated interactions.

The purpose of this paper is:
(i) to investigate the evolution of the spectral properties in the KH
model under increasing frustration of magnetic interactions,
(ii) to recognize the QP behavior in various magnetic phases of the
frustrated KH model, and
(iii) to establish whether the disordered KSL indeed realizes a paradigm
of a Fermi liquid at light doping as suggested in Ref. \cite{Mei12}.
In our study we employ exact diagonalization (ED) of finite periodic 
clusters which has the important advantage that no approximations have 
to be made, which is particularly important as the analytical theory
of hole-motion in the KH model is notoriously difficult and largely 
unexplored. The possible problems with the ED approach are finite size 
effects and the difficulty to extrapolate to the thermodynamic limit 
(TL). As important results, we report the spectral functions of the 
different ordered phases and their respective quasiparticle features.
Moreover we report a totally incoherent spectral weight distribution 
for a hole moving in the Kitaev spin-liquid phase in the strong coupling 
regime, i.e., relevant for the iridate systems. The absence of QPs in 
this case for finite systems clearly suggests that also in the TL there 
are no QPs at strong coupling. 
For this reason we
conclude that then a dilute gas of holes (that are individually not QPs) 
will not turn into a Fermi liquid.

The paper is organized as follows: In the Sect. II we outline the KH model
and discuss the calculation of various correlation functions that are used
to determine the phase diagram of the KH-model. Section III
deals with the motion of holes in the ordered phases of the model
and discusses the different propagators and spectral functions used.
In Sect. IV we address the hole motion in the KSL and
analyse the spectral weight distribution. The latter is discussed
in the intermediate and strong coupling limit. Moreover
the dynamical spin structure factor is analysed for the KSL
both for the Kitaev and the KH model
in order to explore the different scattering channels for
holes. Results are summarized in Sect. V.

\section{The Kitaev-Heisenberg Model}
\subsection{Frustrated spin interactions \\ and exact diagonalization}

\begin{figure}[t!]
\begin{center}
\includegraphics[width=3.8cm]{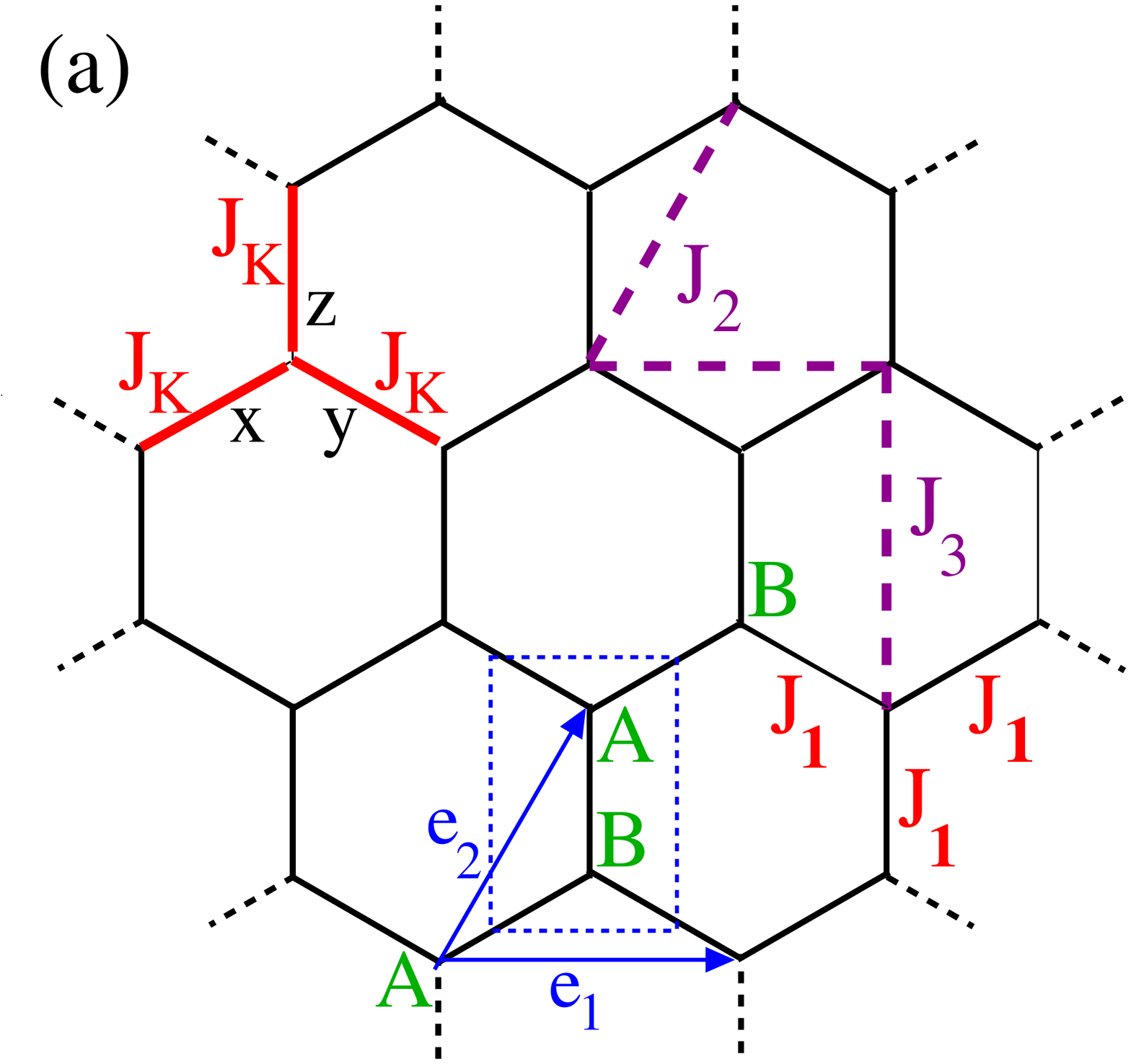} \hskip .0cm
\includegraphics[width=4.0cm]{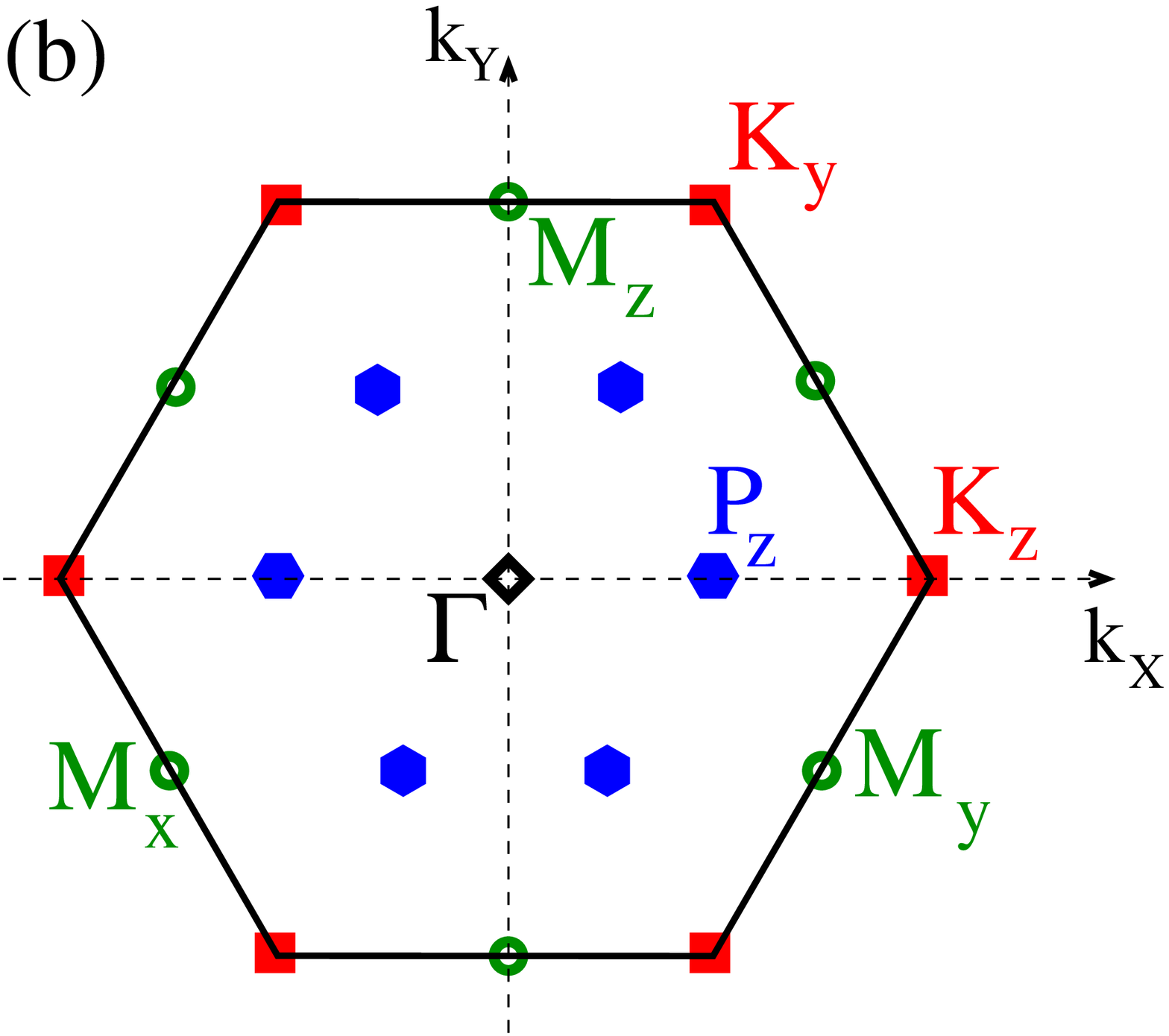} \\
\end{center}
\caption{(Color online)
(a) Periodic cluster of the honeycomb lattice (solid and dashed bonds) 
with $N=24$ sites and the elementary translations $\vec{e}_{1(2)}$ 
that connect atoms $A$ within one sublattice, i.e., connect unit cells
consisting of one atom $A$ and one $B$ each. 
Heisenberg couplings between NN ($J_1$), NNN ($J_2$) and 3NN ($J_3$) in 
${\cal H}_{tJ}$ (\ref{model}) are indicated by solid and dashed lines; 
Kitaev couplings $J_K$ involve a single spin component for each bond
direction, $\gamma\in\{x,y,z\}$.
(b) First BZ with high symmetry $M_\gamma$ and $K_\gamma$ points
(in absence of symmetry breaking these points are equivalent to one
another).}
\label{fig:defs}
\end{figure}

We consider the following $t$-$J$-like KH model ($t>0$),
on the honeycomb lattice [Fig. \ref{fig:defs}(a)]:
\begin{eqnarray}
&{\cal H}_{tJ}&\,\equiv
t\sum_{\langle ij\rangle\sigma} c^{\dagger}_{i\sigma}c^{}_{j\sigma}
 + J_K\sum_{\langle ij\rangle\parallel\gamma} S_i^\gamma S_j^\gamma
 + J_1\sum_{\langle ij\rangle}   \vec{S}_i\!\cdot\!\vec{S}_j \nonumber \\
&+&\!(1-\alpha)\,\Big\{
   J_2\!\!\sum_{\{ij\}\in{\rm NNN}}\!\!\vec{S}_i\!\cdot\!\vec{S}_j
 + J_3\!\!\sum_{\{ij\}\in{\rm 3NN}}\!\!\vec{S}_i\cdot\vec{S}_j\Big\}\,,
\label{model}
\end{eqnarray}
 It consists of the kinetic energy term
$\propto t$ of projected fermions with flavor $\sigma$
\cite{Tro13,Hya12,You12} which move in the restricted space without 
double occupancies as a result of large on-site Coulomb repulsion $U$. 
The spins $S=1/2$ are defined in terms of fermionic creation 
(annihilation) operators $c^{\dagger}_{i\sigma}$ ($c_{i\sigma}$) with 
flavor $\sigma$ at site $i$:
\begin{equation}
S_i^\gamma\equiv \frac12\; c^{\dagger}_{i\sigma}
\tau^{\gamma}_{\sigma\sigma'}c^{}_{i\sigma'},
\label{sigma}
\end{equation}
where $\tau^{\gamma}_{\sigma\sigma'}$ are Pauli matrices,
$\gamma=\{x,y,z\}$.
We emphasize that already the Kitaev terms $\propto J_K$ with different 
Ising spin interactions that depend on bond direction 
introduce strong spin frustration in Eq. \eqref{model}.
In the following we shall assume ferromagnetic (FM) Kitaev ($J_K<0$) 
and AF Heisenberg ($J_1>0$) exchange,
\begin{equation}
J_K\equiv -2J\alpha\,, \hskip .7cm J_1\equiv J(1-\alpha)\,.
\label{JKH}
\end{equation}
Here $J$ is the energy unit and $\alpha\!\in\![0,1]$ is a parameter
that interpolates between the Heisenberg and Kitaev exchange couplings
for NN spins $S=1/2$. The model Eq.~(\ref{model}) includes
NNN $\propto J_2$ and 3NN $\propto J_3$ AF terms as well.

We use exact diagonalization (ED) within the Lanczos algorithm for a
periodic cluster of $N=24$ sites which accomodates all point group
symmetries of the infinite lattice. The momenta corresponding to
allowed symmetry representations are presented in the first Brillouin
zone (BZ) in Fig. \ref{fig:defs}(b).
We introduce $M_z=(0,2\pi/\sqrt{3})$,
$M_{x/y}=(\pi,\pm\pi/\sqrt{3})$, and $\pm P_\gamma$, where
$M_\gamma\perp P_\gamma$ and $2P_\gamma\equiv K_\gamma$.
Note that in absence of symmetry-breaking field, $K_z$ and $-K_y$ are
identical and only two distinct representations exist, $\pm K_z$.

\subsection{Phase diagram}

\begin{figure}[t!]
\begin{center}
\includegraphics[width=6.2cm]{s_J04.eps}
\includegraphics[width=7.0cm]{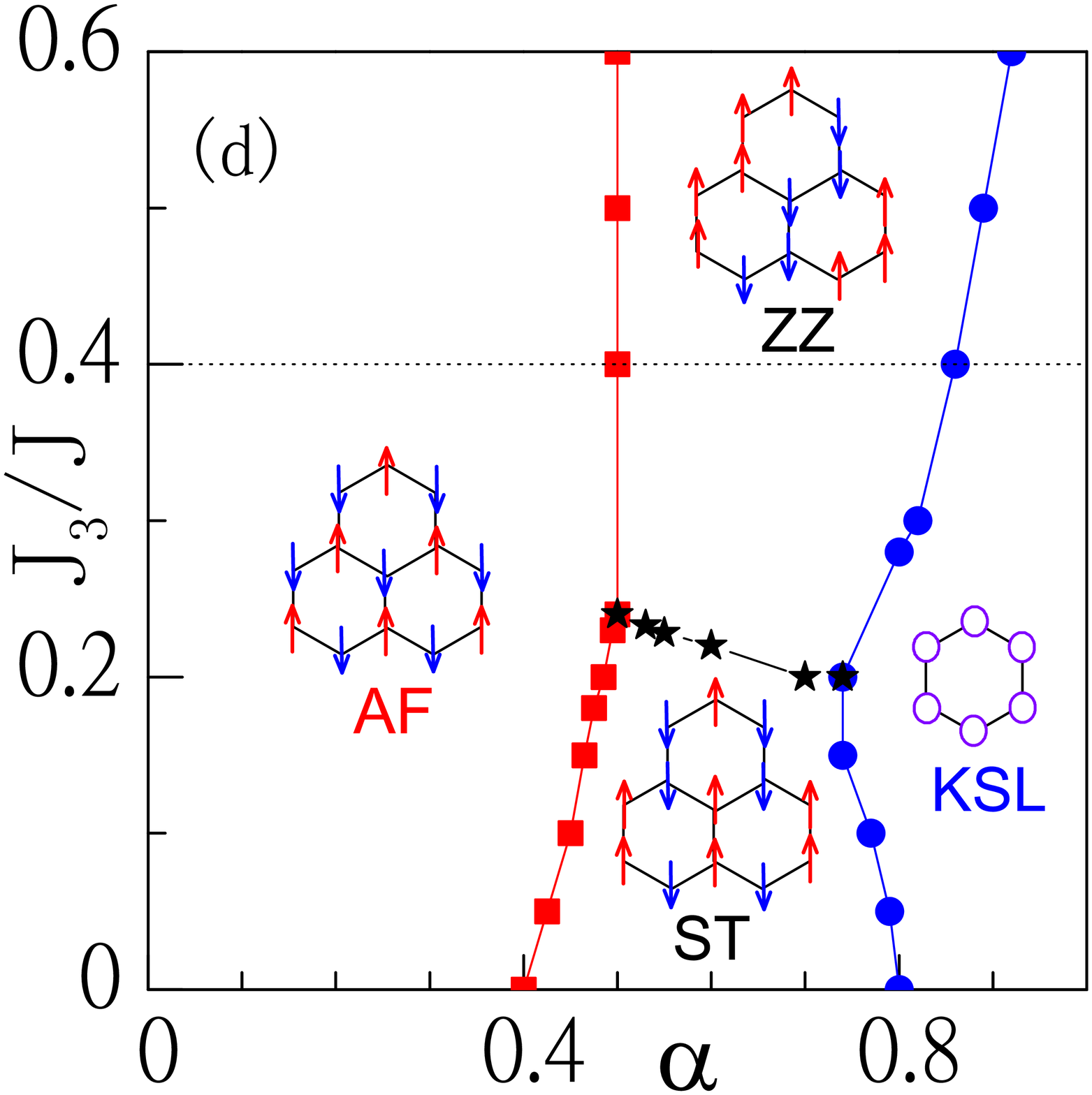}\\
\end{center}
\caption{(Color online)
(a) Order parameters ${\cal S}_{\Phi}$ (\ref{ssf}) representing the 
N\'eel ($\Phi$=AF), the zigzag ($\Phi$=ZZ) and the Kitaev invariant
${\cal L}$ (\ref{w}) obtained for $J_2=0$ and $J_3=0.4J$.
(b)~Phase diagram of the KH model Eq. \eqref{model} in 
$(\alpha,J_3/J)$ plane (points) for $J_2=0$, with AF, ST, ZZ and KSL 
phases 
The insets show types of magnetic order (arrows) or disorder (circles).
}
\label{fig:defsb}
\end{figure}

In the present ED approach of finite systems there is no spontaneous 
symmetry breaking, and the spin components $x$, $y$, and $z$ are 
equivalent. The intrinsic spin order parameter ${\cal S}_{\Phi}$ in 
the ground state can be determined by identifying and  calculating the 
respective correlation functions that reflect the emerging long-range 
order \cite{Kap89,Kom94},
\begin{equation}
{\cal S}_{\Phi}^2\equiv{\frac{12}{N^2}\sum_{i,j=1}^{N/2}
e^{i\vec{k}\cdot\vec{R}_{ij}}
\langle(S^z_{iA}\pm S^z_{iB})(S^z_{jA}\pm S^z_{jB})\rangle}\,,
\label{ssf}
\end{equation}
for each phase $\Phi$, where
$\langle\dots\rangle\equiv\langle 0|\dots|0\rangle$ is the average
over the ground state $|0\rangle$. In this definition we select $"-"$
sign for the spin components $S_{jB}^z$ for the N\'eel AF phase and
$"+"$ sign for the staggered and ZZ phase; $\vec{k}=\Gamma$ for the 
N\'eel ($\Phi={\rm AF}$), $\vec{k}=M_z$ for the ST ($\Phi={\rm ST}$), 
and either $\vec{k}=M_x$ or $\vec{k}=M_y$ for the ZZ ($\Phi={\rm ZZ}$) 
phase (these points are equivalent in this case).
Here $i$ and $j$ label unit cells [Fig. \ref{fig:defs}(a)],
and $\vec{R}_{ij}=\vec{r}_i-\vec{r}_j$.
The order parameter ${\cal S}_{\Phi}$ (\ref{ssf}) is large when spin
correlations are close to the ones expected for a magnetic phase 
$\Phi$; in all other phases it is negligible. Two examples are shown 
in Fig. \ref{fig:defsb}(a):
(i) for $\alpha<0.5$ one finds large ${\cal S}_{\rm AF}\simeq 1$, and
(ii)~for $0.5<\alpha<0.85$ this order parameter drops to
${\cal S}_{\rm AF}\ll 1$ but ${\cal S}_{\rm ZZ}$ increases to
${\cal S}_{\rm ZZ}>0.6$.
Indeed, a transition between the AF and ZZ phase is found here,
see dotted line in Fig. \ref{fig:defsb}(b), while the ST phase is
unstable here and ${\cal S}_{\rm ST}$ is small.

In the KSL phase spin correlations 
$\langle S^{\gamma}_i S^{\gamma}_j\rangle$ vanish beyond NN spins at 
$\alpha=1$ \cite{Kit06,Bas07,Tro11}, and further neighbor correlations 
remain small in the Kitaev liquid regime at $\alpha\neq 1$
\cite{Tro11}. This results in ${\cal S}_{\rm \Phi}\ll 1$ for all 
conventional spin order parameters \cite{SKl}. To identify the KSL we 
introduce here an average of the Kitaev invariant \cite{Kit06} on a 
single hexagon ${\cal C}_6$,
\begin{equation}
{\cal L}\equiv 2^6\left\langle\prod_{i\in{\cal C}_6}
S^\gamma_i\right\rangle,
\label{w}
\end{equation}
where $\gamma$ labels the spin component $S^\gamma_i$ interacting with
a spin at site $j$ along the outgoing bond
$\langle ij\rangle\notin{\cal C}_6$ via $J_KS^\gamma_iS^\gamma_j$.
One finds that ${\cal L}\to 1$ when the ground state of the Kitaev
model is approached at $\alpha\to 1$, see Fig.~\ref{fig:defsb}(a).

The phase diagram in the $J_3$ versus $\alpha$ plane is displayed in 
Fig. \ref{fig:defsb}(b). We recognize the AF phase at small $\alpha$, 
the intermediate ST phase, and at large $\alpha > 0.8$ --- the Kitaev 
liquid phase. At large further neighbor exchange interaction $J_3$ 
the zigzag phase emerges in the intermediate range of $\alpha$. 
Theses phases were identified by the order parameters discussed above,
while the phase boundaries were determined by a different powerful 
tool, namely the study of the fidelity susceptibility \cite{You07,Gu10}, 
i.e., the changing rate of the overlap between ground states at adjacent 
points. Note that the AF$\leftrightarrow$ZZ transition at $\alpha=0.5$ 
follows from symmetry arguments and as such is independent of the 
cluster size.

\section{Carrier propagation in quantum antiferromagnetic phases}

In the following we shall analyze the spectral properties of a hole 
inserted into the ordered 
ground state $|0\rangle$, being the quantum N\'eel 
AF, ST, or ZZ phase. The KSL phase will be explored in detail in the 
subsequent Section. We use here the standard numerical Lanczos 
algorithm which spans efficiently the relevant Krylov space and yields 
spectral functions in form of a continued fraction \cite{Szc90,Gb87}.
The calculation begins with the determination of the ground state 
$|0\rangle$ and the subsequent addition of a hole, that is the 
annihilation of an electron as in a photoemission experiment. 
Therefore 
we consider in the following the hole creation operator, 
$c^{}_{{\vec k}\uparrow}$, in form of a plane wave that includes all 
sites of the honeycomb lattice equally, and alternatively a hole 
creation operator, $d^{}_{{\vec k}\uparrow}$, where holes are created 
only on one sublattice \cite{Sus06}, namely sublattice $A$,
\begin{equation}
c_{{\vec k}\uparrow}=\frac{1}{\sqrt{N}}\sum_i
e^{i\vec{k}\cdot\vec{r}_i} c_{i\uparrow}, \hskip .3cm
d_{{\vec k}\uparrow} \!\equiv\sqrt{\frac{2}{N}} \sum_{i\in A}
e^{i \vec{k}\cdot\vec{r}_i} c_{i\uparrow}.  
\end{equation}
When a hole is created in the ground state $|0\rangle$, the spectral 
functions,
\begin{eqnarray}
\label{Ac}
A_c(\vec k,\omega)\!&=& \frac{1}{\pi}\Im\langle 0|
c^\dagger_{{\vec k}\uparrow}\,\frac{1}{\omega-i\eta+E_0(N)-{\cal H}_{tJ}}\,
c^{}_{{\vec k}\uparrow}|0\rangle,\\
\label{Ad}
A_d(\vec k,\omega)\!&=& \frac{1}{\pi}\Im\langle 0|
d^\dagger_{{\vec k}\uparrow}\,\frac{1}{\omega-i\eta+E_0(N)-{\cal H}_{tJ}}\,
d^{}_{{\vec k}\uparrow}|0\rangle,
\end{eqnarray}
correspond to the physical Green's function $G_c({\vec k},\omega)$
that is measured in ARPES experiments (\ref{Ac}),
or to the sublattice Green's function $G_d({\vec k},\omega)$ (\ref{Ad}). 
In the definition of the spectral functions Eqs. (\ref{Ac}) and 
(\ref{Ad}) excitation energies are measured relative to the ground 
state energy $E_0(N)$ of a Mott insulator with $N$ electrons.

\begin{figure}[t!]
\begin{center}
\includegraphics[width=7.7cm]{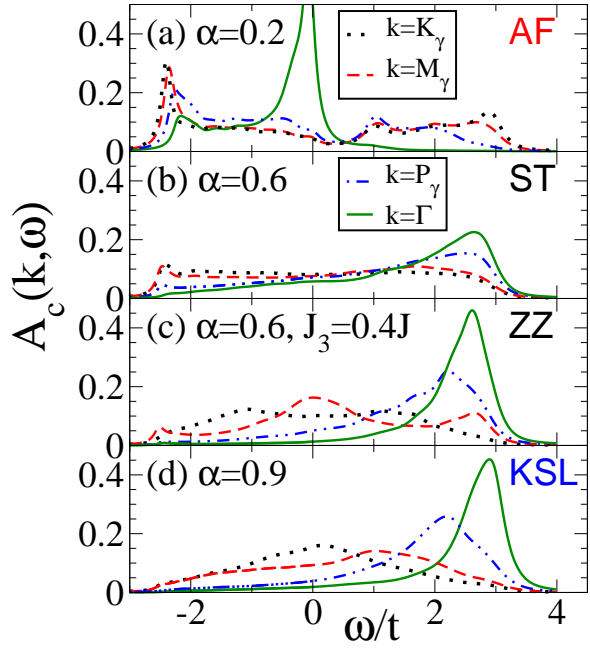}\\
\end{center}
\caption{(Color online)
Spectral function $A_c(\vec k,\omega)$ (\ref{Ac}) for one hole 
excitations \cite{notear} at strong coupling, $t=5 J$,
obtained by ED at four distinct momenta $\vec k$
(solid, dashed, dashed-dotted and dotted lines) for:
(a) the N\'eel phase at $\alpha=0.2$,
(b,c) ST and ZZ phases at $\alpha=0.6$, and
(d) the KSL phase at $\alpha=0.9$.
Parameters: $J_2=0$, $\eta=0.1t$, and
$J_3=0$, except in (c) where $J_3=0.4J$.}
\label{fig:Acw}
\end{figure}

\begin{figure}[t!]
\begin{center}
\includegraphics[width=7.7cm]{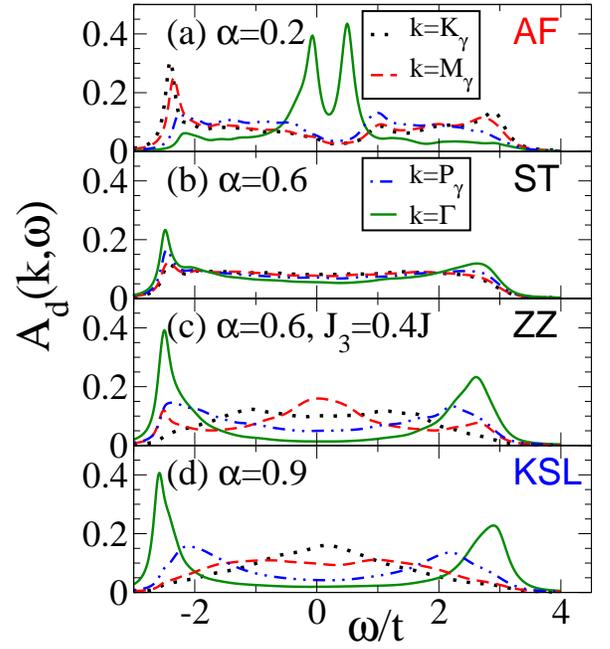}\\
\end{center}
\caption{(Color online)
Spectral function $A_d(\vec k,\omega)$ (\ref{Ad}) for a hole excitation 
created on one sublattice \cite{notear}.
The method, lines and parameters are the same as in Fig. \ref{fig:Acw}.
}
\label{fig:Adw}
\end{figure}

In all phases the spectra $A_c(\vec k,\omega)$ and $A_d(\vec k,\omega)$
shown in Figs. \ref{fig:Acw} and \ref{fig:Adw}, respectively, have the 
total width $W\simeq 6t$ as for free hole motion on the honeycomb 
lattice. For small $\alpha<0.4$ the model \eqref{model} is weakly 
frustrated and $|0\rangle$ is the quantum AF N\'eel state, 
see Fig. \ref{fig:defsb}(b).
Taking $t/J=5$, a value representative for strong coupling ($t>J$)
\cite{noteir}, i.e., when the kinetic energy of a hole is larger than
the energy of a magnetic bond, one finds that the spectral function
$A_c(\vec k,\omega)$ has a QP at low energy and its spectral weight is
large at the $K_{\gamma}$ and much weaker at the $\Gamma$ point, see
Fig. \ref{fig:Acw}(a). In this phase no qualitative differences between
$A_c(\vec k,\omega)$ and $A_d(\vec k,\omega)$ functions are observed, 
see Figs. \ref{fig:Acw}(a) and \ref{fig:Adw}(a),
except near $\omega = 0$ for the $\Gamma$ point.

The sublattice spectral function, $A_d(\vec k,\omega)$, where holes are
injected/removed on the same sublattice, reveals large spectral weight
at low energy. For the ST and ZZ phases the low energy features in
$A_d(\vec k,\omega)$ can indeed be identified as QPs accompanied by
incoherent spectral weight ---
they are more pronounced in the ZZ phase, cf. Figs. \ref{fig:Adw}(b)
and \ref{fig:Adw}(c). Yet, the QP features are suppressed at the
$\Gamma$ and $P$ points in these phases in the physical spectral
function $A_c(\vec k,\omega)$, cf. Figs. \ref{fig:Acw}(b) and
\ref{fig:Acw}(c). As shown before \cite{Tro13}, this is consistent with
the observed absence of QPs in ARPES for Na$_2$IrO$_3$ \cite{Com12}.

\begin{figure}[t!]
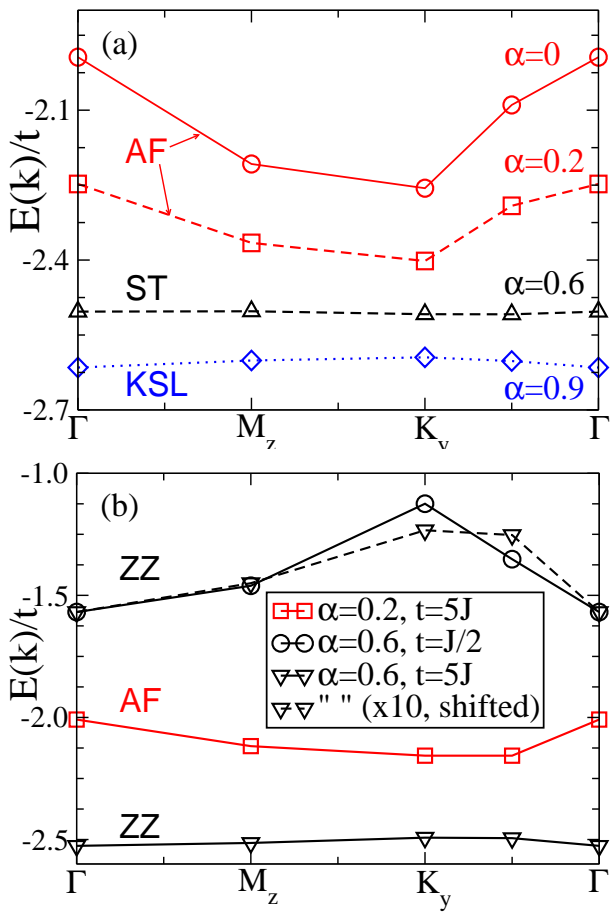

\begin{center}
\includegraphics[width=8cm,height=6cm]{disp1e_N24t20.eps}
\includegraphics[width=8cm,height=6cm]{disp1e_N24J305.eps}\\
\end{center}
\caption{(Color online)
Dispersion of lowest single hole energy obtained for the KH model
Eq.~(\ref{model}) by ED for $t/J=5$, $J_2=0$, and increasing $\alpha$:
(a) $J_3=0$, and (b) $J_3=0.5J$.
Hole energies are respective to the ground state energy $E_0(N)$.}
\label{fig:disp}
\end{figure}

The momentum $\vec k$ dependence of the low energy QPs shown in Fig.
\ref{fig:disp} reveals the strong dependence of hole dispersion on the 
magnetic order. As in cuprates \cite{Mar91}, the QP dispersion in the 
N\'eel AF phase is narrowed from the free (unconstrained) fermionic 
band width $\propto 6t$ by strong correlations and is determined by the 
magnetic exchange $\propto J_1$. The dispersion has a minimum (maximum) 
at the $K$ ($\Gamma$) point and is further reduced when frustration of 
magnetic exchange increases from $\alpha=0$ to $\alpha=0.2$ 
[Fig.~\ref{fig:disp}(a)]. At finite $J_3$ the hole energy decreases at 
the $P$ point, but otherwise the dispersions at $\alpha=0.2$
are quite similar, cf. Figs. \ref{fig:disp}(a) and \ref{fig:disp}(b).

In contrast, the dispersion is absent in the ST phase,
see Fig.~\ref{fig:disp}(a),
as coherent hole propagation is hindered here due to the alternating AF 
and FM bonds. Instead, for FM chains in the ZZ phase the dispersion 
appears reversed with respect to that found for the N\'eel phase 
--- now a minimum (maximum) is at the $\Gamma$ ($K$) point. While this
dispersion decreases from weak ($t=J/2$) to strong ($t=5J$) coupling,
its shape remains the same [dashed line in Fig. \ref{fig:disp}(b)].

At first glance one might conclude from the spectral functions for the 
KSL, displayed in Figs. \ref{fig:Acw}(d) and \ref{fig:Adw}(d), that 
hole propagation in the KSL is similar to that in the ZZ phase 
[Figs. \ref{fig:disp}(a) and \ref{fig:disp}(b)].
In this case the lowest excitation energy is found at the $\Gamma$ 
point. Again, one finds a distinct peak in $A_d(\Gamma,\omega)$ which 
is absent in $A_c(\Gamma,\omega)$, suggesting also here a hidden QP.
As shall see below that the fine structure of 
the low energy peaks in the spectral function shown in Fig. 
\ref{fig:Adw}(d) does not represent well defined QPs
in the case of the KSL.

\section{Carrier propagation in the Kitaev spin-liquid phase}

\subsection{Spectral weight distribution}

\begin{figure}[t!]
\begin{center}
\includegraphics[width=8.0cm]{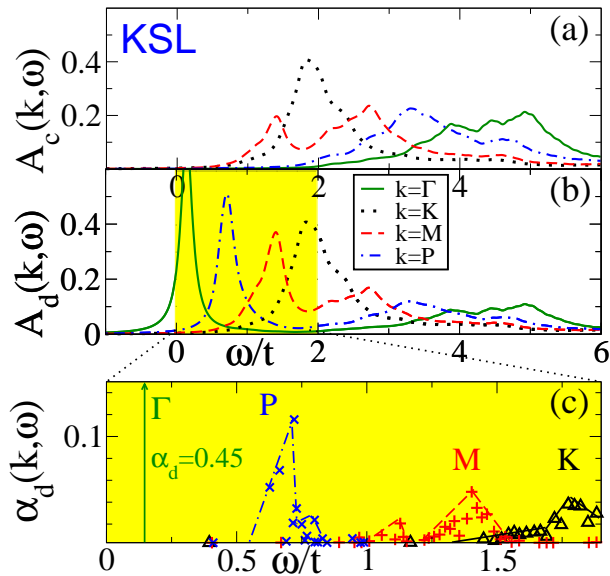}\\
\end{center}
\caption{(Color online)
Spectral functions for the KSL phase obtained by ED in the
intermediate coupling regime $t/J=0.25$ at four nonequivalent momenta
${\vec k}$ (solid, dashed, dashed-dotted and dotted lines) with
$\eta=0.1t$:
(a) full spectral function $A_c(\vec k,\omega)$, and
(b) sublattice function $A_d(\vec k,\omega)$.
Panel (c) shows the spectral weights $\alpha_d({\vec k},\omega)$
(\ref{alpha}) (symbols) larger than $3\times 10^{-3}$ at low energy
(shaded) for different ${\vec k}$ values; dashed lines are guides to
the eye. Parameters: $\alpha=0.9$, $J_2=J_3=0$.
}
\label{fig:ksl}
\end{figure}

Next we show that the KSL phase is manifestly different from all the
ordered phases discussed so far. We address the nature of low energy
states by analyzing first the intermediate coupling regime of 
$t/J=0.25$. Again one finds distinct low energy peaks in 
$A_d(\Gamma,\omega)$ \cite{note1}, see Fig. \ref{fig:ksl}(b),
missing in $A_c({\vec k},\omega)$ [Fig. \ref{fig:ksl}(a)].
In contrast, both spectral functions are rigorously identical at the 
$K$ point, cf. $A_c(K,\omega)$ and $A_d(K,\omega)$ in
Figs. \ref{fig:ksl}(a) and \ref{fig:ksl}(b). We note that for the
honeycomb lattice $K$ corresponds to the Dirac point which is the
degeneracy point of noninteracting electrons \cite{Gra09}.

A surprise comes when the fine structure of the sublattice 
spectral function is analyzed in absence of spectral broadening 
[at $\eta=0$ in Eq. (\ref{Ad})],
\begin{equation}
A_d^{(0)}(\vec{k},\omega)=\lim_{\eta\rightarrow 0}A_d(\vec{k},\omega),  
\end{equation}
which may be rewritten using spectral weights:
\begin{equation}
\label{Ada}
A_d^{(0)}(\vec{k},\omega)=
\sum_n\alpha_d({\vec k},\omega_n)\delta(\omega-\omega_n)\,.
\end{equation}
In the following we shall see the advantage of studying the 
single-particle propagation directly in terms of the 
{\it spectral weight distribution},
\begin{equation}
\label{alpha}
\alpha_d({\vec k},\omega_n)=
|\langle\psi_{\vec{k},n}{(N-1)}|d_{\vec{k},\uparrow}|0\rangle|^2,
\end{equation}
at excitation energies 
\label{omega}
\begin{equation}
\omega_n=E_{\vec{k},n}{(N-1)}-E_0(N),
\end{equation}
and $|\psi_{\vec{k},n}(N-1)\rangle$ is an excited state in the space 
with one extra hole and total momentum $\vec{k}$ that contributes with 
a finite spectral weight. 
We stress, that $\alpha_d({\vec k},\omega_n)$
is in mathematical terms a distribution and not a function, 
although in the TL it may contain parts that can be represented 
by continuous curves, in some cases.

When looking at the spectral weight distribution  in Fig. 
\ref{fig:ksl}(c) we recognize a robust QP, that is, a well separated 
bound state at lowest energy, only at the $\Gamma$ point with
large spectral weight $\alpha_d(\Gamma,\omega_0)\simeq 0.45$. 
At {\it all other\/} ${\vec k}$ points no bound states exist but instead 
rather continuous distributions of weights $\alpha_d({\vec k},\omega_n)$ 
is found in the range of the lowest eigenvalues $\omega_n$ (\ref{omega}). 
From the spectral weight distribution 
displayed in Fig. \ref{fig:ksl}(c) it is clear that for 
example at the $P$ point the spectra are represented by a superposition 
of several many-body states in the $(N-1)$-particle sector with 
slightly different energies and there is no dominant pole that could be 
considered as a QP state. On the other hand, simply by looking at the 
spectral function $A_d(\vec{k},\omega)$ in Fig. \ref{fig:ksl}(b) one 
may easily overlook the breakdown of the QP picture. We stress that in 
all cases there is substantial incoherent spectral weight at higher 
energies that extends to the upper edge of the spectrum at $6t$.

To understand this striking difference of the character of the spectra 
at low energy at $\Gamma$ and $P,M,K$, respectively, we need to 
investigate the spin excitations in the Kitaev liquid regime. Finally 
it is the scattering of carriers from spin excitations that determines 
whether they can propagate as QPs or whether they are completely 
overdamped.

\subsection{Spin excitations in Kitaev spin liquid}

Our aim here is to explore the spin excitations of the KH
model in the KSL regime. However, our discussion will be more 
transparent when we first focus on the pure Kitaev model at $\alpha=1$ 
and subsequently analyze by numerical simulation the changes of the 
spin structure factor for the KH case in the KSL regime 
($0.8 < \alpha < 1.0$).

We shall employ here the usual representation of the Kitaev model in 
terms of $\sigma^{\gamma}_i=2 S^{\gamma}_i$ spin operators
\begin{equation}
\label{HK}
H_K=-\frac{1}{2}J\sum_{\langle ij\rangle\parallel\gamma} 
\sigma_i^\gamma \sigma_j^\gamma,
\end{equation}
where $J_K=-2 J \alpha$ and $\alpha=1$, according to Eq. \eqref{model}.
Spin excitations are most easily understood by transforming the Kitaev 
model into the Majorana representation \cite{Kit06}. Kitaev introduced 
a representation for the spin algebra
$[\sigma_i^a,\sigma_j^b]=i \epsilon_{abc}\sigma_i^c\delta_{ij}$
in terms of four Majorana operators $\eta_i^{\gamma}$, $\gamma=0,x,y,z$, 
per lattice site which obey the anticommutation relations,
$\{\eta^{\gamma},\eta^{\gamma'}\}=2 \delta_{\gamma,\gamma'}$.

Here each spin operator component is expressed by a product of two 
Majorana fermions,
\begin{equation}
\label{major}
\sigma_i^\alpha=i \eta_i^0\eta_i^\alpha, \hskip .5cm \alpha=x,y,z,
\end{equation}
where  $\eta_i^\alpha$ is associated with the bond direction and 
$\eta_i^0=\eta_i$, with the respective vertex (at site $i$). 
We suppress the upper index in $\eta_i^0$ further on. 
Using this representation one can write the Hamiltonian as follows,
\begin{equation}
\label{HKa}
H_K=-\frac{1}{2}J\sum_{\langle ij\rangle_\gamma}i 
\eta_i u_{\langle ij\rangle}^{\gamma} \eta_j, 
\end{equation}
where $u_{\langle ij\rangle}^\alpha$ are the bond operators,
\begin{equation}
u_{\langle ij\rangle}^\alpha=i\eta_i^a\eta_j^\alpha.
\label{bondo}
\end{equation}
To take care of the fermionic property 
$u_{\langle ij\rangle}^\alpha=-u_{\langle ji\rangle}^\alpha$ we
adopt a notation where $i(j)$ is on the $A(B)$ sublattice, respectively.
The important point is now, that the so-defined bond operators
\eqref{bondo} commute with the Hamiltonian,
\begin{equation}
\label{u}
[H_K,u_{\langle ij\rangle}^\alpha]=0.
\end{equation}
Hence they are conserved quantities and due to their definition 
as products of two Majoranas, see Eq. \eqref{bondo}, these 
operators can only take the values $u_{\langle ij\rangle}^\alpha=\pm 1$.

In the ground state sector all $u_{\langle ij\rangle}^\alpha$ have the 
same sign on all bonds, either $+1$ or $-1$. A change of a bond 
variable leads to a {\it vortex pair excitation} with a gap 
$\Delta_P\simeq 0.26|J_K|$.
In addition there is a second class of spin excitations which result
from the motion of the $\eta$ Majorana particles 
described by the Hamiltonian in Eq. \eqref{HKa}. Here 
we concentrate on the symmetric case, with equal exchange constants
along three directions in the honeycomb lattice, $J_x=J_y=J_z$;
in this case the {\it Majorana excitations are gapless} \cite{Kit06}.

\begin{figure}[t!]
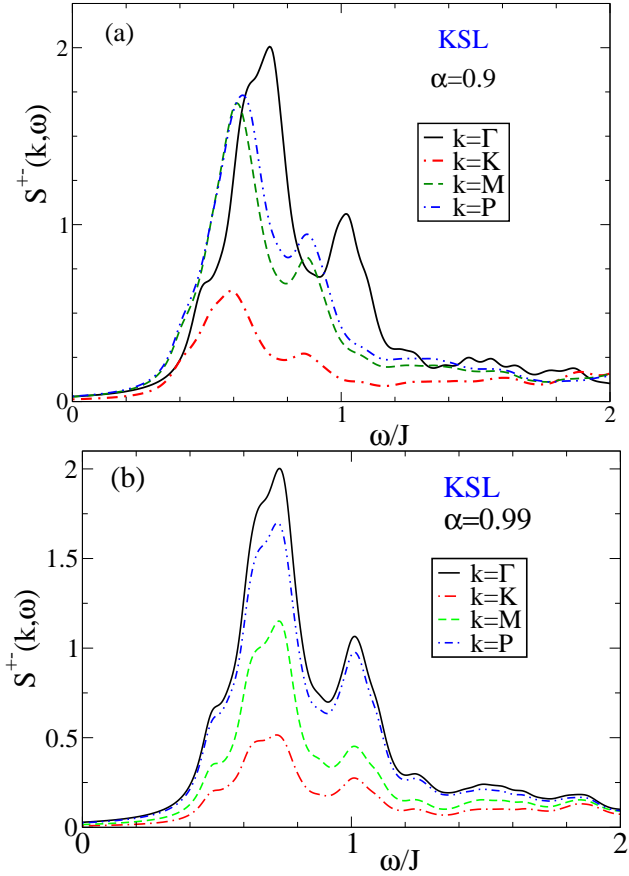

\begin{centering}
\includegraphics[clip,width=8cm]{skw_a.eps}
\vspace{0.3cm}
\includegraphics[clip,width=8.3cm]{skw_b.eps}
\par\end{centering}
\caption{(Color online)
Dynamic spin-structure factor $S^{+-}({\vec k},\omega)$  \eqref{spins} 
for the undoped KSL in the case of the Kitaev-Heisenberg model 
Eq.~\eqref{model} at four distinct momenta, for:
(a) $\alpha=0.9$, and
(b) $\alpha=0.99$.
The curves are obtained by exact diagonalization of a 24-site cluster
with periodic boundary conditions and are smoothened using a Lorentzian
broadening $\eta=0.05J$.}
\label{fig:skw}
\end{figure}

Next we shall investigate the dynamic spin-structure factor in the 
Kitaev limit $\alpha\rightarrow 1$ by exact diagonalization and explore 
the changes in the Kitaev-liquid regime of the KH 
model for $\alpha\neq 1$, which cannot be solved analytically.
This is demonstrated by a calculation of the dynamic spin-structure
factor $S^{+-}({\vec k},\omega)$ defined as follows \cite{Toh95},
\begin{equation}
\label{spins}
S^{+-}({\vec k},\omega)=\frac{1}{N}\sum_{\nu}
|\langle{\nu}|S^+_{\vec k}|0\rangle|^2 \delta(\omega-E_{\nu}+E_0),
\end{equation}
where
$S^{\gamma}_{\vec k}=\sum_{\vec r}e^{i{\vec k}\cdot{\vec r}}
S_{\vec r}^{\gamma}$, and $\gamma=+,-$ denotes spin raising (lowering)
operators, respectively. Here $|0\rangle$ and $|\nu\rangle$ are the
ground and excited state, with energies $E_0$ and $E_{\nu}$, 
respectively. The spin quantization axis is chosen parallel to the
$z$-th spin axis of the Kitaev term for convenience.

Figure \ref{fig:skw}(a) shows $S({\vec k},\omega)$ for the KH model 
Eq. (1) in the KSL regime at $\alpha=0.9$ for four different $\vec{k}$ 
points: $\Gamma$, $K$, $M$ and $P$.
The numerical calculations were performed for a 24-site cluster
with periodic boundary conditions.
The main weight of the spin structure factor is concentrated between
the vortex type spin gap at $\Delta_s\sim 0.4J$ and $\omega\sim 1.2J$. 
We observe a moderate momentum dependence with the vortex spin gap at
$\Gamma$, $\Delta_s\sim 0.46 J$ slightly larger than at $K$, $M$, and 
$P$, where we find $\Delta_s\sim 0.40 J$. Moreover, we can conclude 
that the dispersion of the gap and of the peak structures is due to the 
Heisenberg terms in the Hamiltonian and disappears when one approaches 
the Kitaev limit \cite{Kno13}, as seen in Fig.~\ref{fig:skw}(b).

Thus we find here that the main contributions to the spin-response as 
measured by dynamical spin-structure factor \eqref{spins} come from
the gapped vortex excitations. Furthermore the 
classification of spin excitations obtained for the pure Kitaev model
as well as the spin gap due to vortex excitations
appears still relevant for the KSL phase of the KH 
model Eq. (1) at $\alpha=0.9$.

\subsection{Hole motion in the Kitaev spin liquid: \\ 
Intermediate coupling}

Our central argument why QPs at finite momentum are destroyed at 
{\it intermediate coupling}, as shown in Fig. \ref{fig:ksl}(c), 
will be outlined next. Our explanation
rests on the fact that there are two distinct 
types of elementary spin excitations from which the holes can scatter 
in the KSL. These excitations can be classified according to the exact 
solution given by Kitaev \cite{Kit06} as:
(i) gapped vortex spin excitations with minimal gap $\Delta_s$, and
(ii) gapless Majorana excitations.

We consider as intermediate coupling regime the range $0.1J<t<J$, 
where the kinetic energy of holes is comparable with the magnetic 
energy on the bonds, and not much larger as in the strong coupling 
regime, $J\ll t$. An important aspect of the spectral functions at 
intermediate coupling, as in the case of Fig. \ref{fig:ksl}(c) which 
was calculated for $J=4t$, is the large size of the spin gap,  
$\Delta\simeq 0.4J\simeq 1.6t$ in units of the $t$-scale. As the 
excitation energies $\omega_n$ of the low energy states of holes
at $P$, $M$ and $K$ relative to the bottom of the band at the $\Gamma$
point $\omega_0(\Gamma)$, are much smaller than the vortex spin gap 
$\Delta\simeq 1.6t$ at $J=4t$, no decay via vortex excitations 
\cite{Kit06} can occur. Therefore, we conclude that the destruction 
of QPs is due to the scattering of {\it gapless\/} Majorana fermion 
excitations, and that Fig. \ref{fig:ksl}(c) has to be seen as a 
fingerprint of fractionalization of electrons into holons and gapless 
Majorana fermions of the Kitaev model \cite{Kit06}. Because of the 
dominant role of scattering from gapless spin-excitations
we expect that in larger clusters with a denser $\vec{k}$-mesh, hole 
pockets near the $\Gamma$ point are not protected against strong
scattering, and in consequence the KSL does not turn into a Fermi
liquid at low doping.

\begin{figure}[t!]
\begin{centering}
\includegraphics[clip,width=8cm]{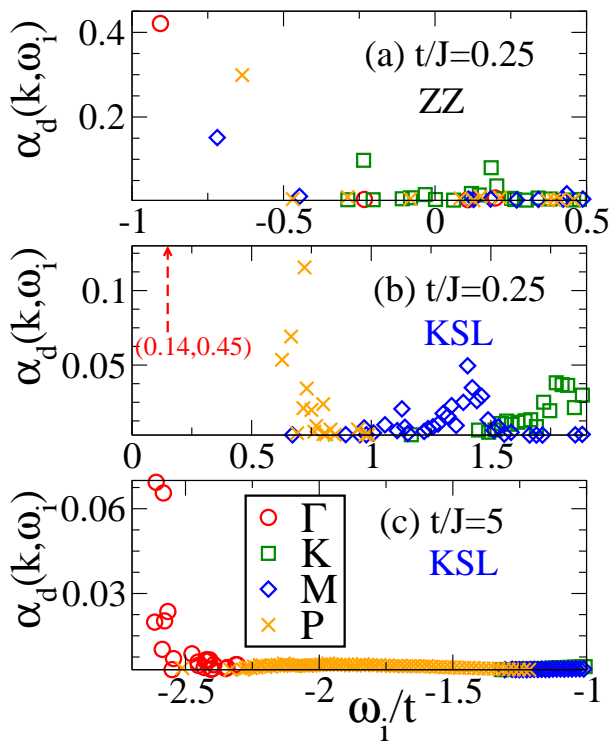}
\par\end{centering}
\caption{(Color online)
Spectral weight distribution $\alpha_d(k,\omega)$, as obtained for:
(a) the ZZ  phase and
(b) the KSL phase, both at intermediate coupling $t=0.25J$, and
(c) the KSL phase at strong coupling $t=5J$.
Parameters: (a) $\alpha=0.4$ and $J_3=0.5$;
(b) and (c) $\alpha=0.9$ and $J_3=0$ in Eq. (1).
}
\label{fig:qp}
\end{figure}

Before moving to the strong coupling regime $t\gg J$ we shall 
highlight the striking difference of the spectral weight distribution 
$\alpha_d(\vec{k},\omega_n)$ \eqref{alpha}
at low energy in the ZZ and the KSL phase, respectively, at $t=0.25J$.
In Fig. \ref{fig:qp}(a) the spectral weight distribution 
$\alpha_d(\vec{k},\omega_n)$ for the ZZ phase shows well defined QP 
bound states. Although the spectral weights of these QPs are 
significantly reduced, the bound states are well separated from the 
continuum (of incoherent states) at higher energy, except for 
$\vec{k}=K$ where this binding energy is much weaker.
In the KSL phase Fig. \ref{fig:qp}(b), on the other hand, 
a well defined bound state is seen only at the $\Gamma$ point.
Because of the size 
of the vortex gap $\Delta_s\simeq 1.6t$ the absence of QPs in Fig. 
\ref{fig:qp}(b) at $P$, $M$ and $K$ is a smoking gun for the 
important role played by scattering by gapless Majorana excitations. 
This is consistent with the spectral shape of
$\alpha(\vec{k},\omega)$ which is reminiscent of a continuum, 
that may result from 
a convolution of holons and gapless Majorana fermions.

\subsection{Hole motion in the Kitaev spin liquid: \\ Strong coupling}

We turn now to the {\it strong coupling\/} 
regime. In the strong coupling case of $t/J=5$, i.e., relevant for the 
iridates \cite{Com12} and displayed in Fig. \ref{fig:qp}(c),
the vortex spin gap becomes small (in units of $t$), i.e.,
$\Delta_s=0.08 $t. 
Thus the strong coupling result in Fig. \ref{fig:qp}(c) highlights the 
effect of the additional vortex excitations which form a new decay 
channel and damp the excitations at the $K$, $M$, and $P$ points even
further. Compared to the result at intermediate coupling at $t=J/4$ 
[Fig. \ref{fig:qp}(b)] where the vortex gap is $\Delta_s=0.4J=1.6t$, 
at strong coupling one finds a spin gap to the vortex excitations 
which is 20 times smaller, that is $\Delta_s=0.08t$.

Furthermore, it may be instructive to go back to the spectral function 
$A_d(k,\omega)$ in Fig. \ref{fig:Adw}(d) which basically 
contains the same information for the strong coupling case ($t=5J$) 
of the KSL as the spectral weight distribution shown in 
Fig. \ref{fig:qp}(c). The fine structure seen in  
$\alpha_d(\vec{k},\omega_n)$ can only be resolved in $A_d(k,\omega)$ 
when the resolution parameter $\eta$ is taken small enough. For ARPES 
experiments, where actually $A_c(k,\omega)$ is measured, this implies 
that a sufficient momentum and energy resolution is required.

We conclude, that the peak at $\Gamma$, which appeared as a separate 
bound state at intermediate coupling in Fig. \ref{fig:qp}(b), appears 
now at strong coupling, see Fig. \ref{fig:qp}(c), rather as the edge of 
the continuum than as an isolated bound state. Moreover, when a single 
hole does not propagate as a QP, then a dilute gas of holes will not 
form a Fermi liquid. Therefore
we stress, that the absence of a well defined separate bound state is 
a {\it new\/} qualitative feature which speaks against QPs and Fermi 
liquid behavior at low doping, and this time the argument emerges
from a diagnosis at the $\Gamma$ point itself!

\section{Summary}
\label{sec:summa}

We have shown that hole propagation is modified in a
remarkable way as increasing Kitaev interactions drive the system from 
the N\'eel order via other ordered antiferromagnetic phases towards 
the Kitaev spin liquid. Quasiparticles are found in the N\'eel phase,
whereas coherent hole propagation is hindered in stripe and zigzag
phases, where hidden quasiparticles with weak dispersion result
from coexisting ferromagnetic and antiferromagnetic bonds.
As the most unexpected result, in the Kitaev liquid phase we have
found unprecedented spectral weight distribution at low energy that
signals the absence of quasiparticles, both at intermediate and strong
coupling. Thus, it appears clearly that carrier motion in the lightly
doped Kitaev liquid is non-Fermi liquid like.

The above conclusion follows from the short-range nature of spin 
correlations in the Kitaev spin liquid \cite{Bas07}. Unlike in a
quantum antiferromagnet on a square \cite{Mar91} or honeycomb 
\cite{Sus06} lattice where spin-flip processes couple to a moving 
hole and generate new energy scale for coherent hole propagation, 
the Kitaev spin-liquid phase is characterized by Ising-like nearest 
neighbor spin correlations of one spin component. Such correlations 
are insufficient to generate coherent hole propagation. They are 
well captured by the present cluster size of $N=24$ sites, and 
qualitative changes of the described scenario are therefore 
unexpected when the cluster size is increased.

Summarizing, we have given clear arguments that shed serious doubts
on the claim from a slave-boson approximation, namely that the low
doped Kitaev spin-liquid phase is a Fermi liquid \cite{Mei12}. The 
first of these arguments rests on the exact solution for the spin 
excitations in the Kitaev model in the intermediate coupling regime, 
and on our finding that {\it gapless} Majorana excitations are 
responsible for the absence of QPs away from the $\Gamma$ point. 
From the gaplessness we conclude that
also states in the closer vicinity of $\Gamma$ are not protected.

Our second argument emerges in the strong coupling case from the result
for the spectral distribution $\alpha_d(\vec{k},\omega)$ at the $\Gamma$
point itself, which in this case has no similarity to a QP but appears
rather as the edge of a continuum. We consider this as evidence that at
strong coupling there are no QPs in the single hole case near the 
$\Gamma$ point, and from this finding we can safely conclude that
Fermi liquid behavior is absent in the low doping regime.

\acknowledgments

We thank Bruce Normand for valuable advice, as well as Mona Berciu,
George Jackeli and Roser Valent\'i for insightful discussions.
A.M.O. acknowledges support by the Polish National Science
Center (NCN) under Project No. 2012/04/A/ST3/00331.
W.-L.Y. acknowledges support by the Natural Science Foundation 
of Jiangsu Province of China under Grant No. BK20141190.


\begin{thebibliography}{99}
\bibitem{SKl} The relatively large value ${\cal S}_{\rm ZZ}\simeq 0.44$
              seen for $\alpha \rightarrow 1$ is due to the smallness
              of the cluster and consistent with spin correlations
              calculated in Ref. \cite{Bas07}.
\bibitem{note1} Spectral function $A_d(\Gamma,\omega)$ is similar at low
                $\omega$ for $t=J$.
\bibitem{notear} In our definition of $A_c({\vec k},\omega)$ Eq.
                 (\ref{Ac}), we plot increasing ARPES excitation
                 energies from left to right.
\end{thebibliography}
\end{document}